\documentclass[pra,aps]{revtex4}

\usepackage{amsmath}
\usepackage{ulem}
\usepackage{bm}
\usepackage{graphics}
\usepackage[dvips]{graphicx,color}
\usepackage{dcolumn}

\begin{document}
\title{Equilibrium Properties of a Trapped Dipolar Fermion at Finite Temperatures}
\author{Yuki Endo$^1$, Takahiko Miyakawa$^2$ and Tetsuro Nikuni$^1$}
\affiliation{$^1$Department of Physics, Faculty of Science, Tokyo University of Science, 
1-3 Kagurazaka, Shinjuku-ku, Tokyo 162-8601, Japan\\
$^2$Faculty of Education, Aichi University of Education, 1 Hirosawa, Igaya-cho, Kariya, Aichi 448-8542, Japan}
\date{\today}

\begin{abstract}
We study the equilibrium properties of a dipolar Fermi gas at finite temperatures. We introduce a variational ansatz for the phase-space distribution function that can describe the deformation in both real and momentum space. The effect of dipole--dipole interactions on the thermal equilibrium is discussed with particular emphasis on the deformation in momentum space. We examine the stability of the system by varying the temperature, trap aspect ratio, and the dipole moment. In addition, we discuss how the deformation in both real and momentum space can be observed in the high-temperature regime.
\end{abstract}

\maketitle
Interest in dipolar gases has been growing since Bose--Einstein condensates (BECs) of ${\rm ^{52}Cr}$ atoms, which have large magnetic dipole moments, were experimentally observed~\cite{Griesmaier2005,Stuhler2005,Giovanazzi2006}. The anisotropic and long-range nature of the dipole--dipole interaction confers rich properties to both the equilibria and dynamics of dipolar gases. There have been a number of theoretical studies of dipolar BECs, which have investigated their ground states~\cite{Santos2000,Yi2000}, collective oscillations~\cite{Yi2001,Goral2002_2}, and their properties in optical lattice potentials~\cite{Goral2002,Danshita2008}. The ground state~\cite{Goral2001,Miyakawa2008} and stability~\cite{Miyakawa2008,Zhang2009} of dipolar Fermi gases have been studied. In addition, there have been theoretical studies of expansion~\cite{He2008,Nishimura2009}, collective oscillations~\cite{Sogo2009}, and the superfluid phase~\cite{Baranov2002,Baranov2004,Bruun2008,Zhao2009}. On the other hand, several experiments proceed energetically toward the realization of creating heteronuclear polar molecules whose large electric dipole moments result in strong dipole--dipole interactions. Rydberg atoms have also been attracting attention due to their large electric dipole moments.

Most studies of dipolar Fermi gases have focused on the zero-temperature properties of these gases and there have been few theoretical studies on finite-temperature gases~\cite{Zhao2009,Zhang2010,Kestner2010}. Despite many groups conducting experiments, no groups have succeeded in cooling polar molecules to the quantum-degenerate regime~\cite{Ni2008}. It is thus important to investigate the temperature range in which the dipole--dipole interaction has appreciable effects. For this reason, we concentrate on the properties of dipolar Fermi gases at finite temperatures.

Unlike dipolar BECs, dipolar Fermi gases do not interact via {\it s}-wave collisions and they have both Hartree direct and Fock exchange energies of the dipole--dipole interaction in the mean-field description, which reflects the antisymmetric many-body wave function. Miyakawa {\it et al.} introduced a variational Wigner function to examine the ground state of the system at zero temperature and they demonstrated that the anisotropic nature of the interaction causes Fermi surface deformation through the Fock exchange energy~\cite{Miyakawa2008}. In the present paper, we generalize this variational method to finite temperatures and discuss the effect of temperature on the equilibrium properties. In particular, we focus on the deformation of the distributions in real and momentum space and investigate the temperature dependence of the instability of dipolar Fermi gases against collapse.

We consider trapped dipolar fermions. The dipoles are assumed to be polarized along the ${\it z}$ axis due to an external electric field. In the second quantized form, the Hamiltonian for this system is given by
\begin{eqnarray}
\hat{H}&=&\int d\textbf{r}\hat{\Psi}^\dagger\left(\textbf{r}\right)\left[-\frac{\hbar^2}{2m}\nabla_{\textbf{r}}^2+V_{trap}\left(\textbf{r}\right)\right]\hat{\Psi}\left(\textbf{r}\right)\nonumber \\
	&+&\frac{1}{2}\int d\textbf{r}\int d\textbf{r}^\prime\hat{\Psi}^\dagger\left(\textbf{r}\right)\hat{\Psi}^\dagger\left(\textbf{r}^\prime\right)V_{dd}\left(\textbf{r}-\textbf{r}^\prime\right)\hat{\Psi}\left(\textbf{r}^\prime\right)\hat{\Psi}\left(\textbf{r}\right),\label{Hamiltonian}
\end{eqnarray}
where $\hat\Psi\left(\textbf{r}\right)$ is the Fermi field operator, and the hat indicates a second quantized operator. The first term of Eq. (\ref{Hamiltonian}) describes the Hamiltonian of a single particle in a harmonic trap $V_{trap}\left(\textbf{r}\right)=\left(m/2\right)\left[ \omega_\rho^2\left(x^2+y^2\right)+\omega_z^2 z^2 \right]$, where ${\it m}$ is the particle mass. The second term describes the two-body interaction Hamiltonian of the dipole--dipole force, where $V_{dd}\left(\textbf{r}\right)=\left(d^2/r^3\right)\left(1-3z^2/r^2\right)$. 
$d$ is the coefficient of the dipole--dipole interaction and $d^2=p^2/4\pi\epsilon_0$, where $p$ is magnitude of the electric dipole moment~\cite{Miyakawa2008,Sogo2009} and $\epsilon_0$ is the electric permittivity of a vacuum.
We introduce the Wigner distribution function:
\begin{eqnarray}
{W}\left( \textbf{p},\textbf{r} \right) = \int d \textbf{r}^\prime e^{-i\textbf{p} \cdot \textbf{r}^\prime / \hbar }
          \langle\hat \Psi ^\dagger\left( \textbf{r} - \textbf{r}^\prime /2 \right)\hat \Psi\left( \textbf{r} + \textbf{r}^\prime /2 \right)\rangle.\label{Wigner}
\end{eqnarray}
The density distributions in real and momentum space in terms of ${W}\left( \textbf{p},\textbf{r} \right)$
are given by $n\left(\textbf{p}\right)= \int d\textbf{r}\ W\left(\textbf{p},\textbf{r}\right)$
and $n\left(\textbf{r}\right)=\left(2\pi\hbar\right)^{-3}\int d\textbf{p}\ W\left(\textbf{p},\textbf{r}\right)$, respectively.

We consider the thermal equilibrium of dipolar fermions trapped in a harmonic potential.
The system is not globally stable when a dipole--dipole interaction is present because the interaction is partially attractive and causes collapse of the gas. However, a metastable state exists at finite temperatures under certain conditions.
To find this metastable state, we introduce a variational Wigner distribution function that is analogous to the one introduced in Ref.~\cite{Miyakawa2008}. We assume the Maxwell--Boltzmann regime at relatively high temperatures, namely
\begin{eqnarray}
W\left(\textbf{p},\textbf{r}\right)=\exp\left\{-\left[\frac{\theta^2}{2m}\left(\frac{p_\rho^2}{\alpha}+\alpha^2p_z^2\right)+\frac{\lambda^2m\omega^2}{2}\left(\beta\rho^2+\frac{z^2}{\beta^2}\right)-\mu_0\right]/k_BT\right\},\label{Wigner2}
\end{eqnarray}
where $\rho^2\equiv x^2+y^2$, $p_\rho^2\equiv p_x^2+p_y^2$, and $\omega\equiv\left(\omega_\rho^2\omega_z\right)^{1/3}$. 
Here, the positive parameters ${\alpha}$ and $\beta$ represent deformations of density distributions in momentum and real space respectively, and $\lambda$ describes compression of the dipolar gas, as defined in Ref.~\cite{Miyakawa2008}. In addition to these parameters, we introduce a new variational parameter $\theta$ that characterizes compression in momentum space. In addition, the chemical potential is determined by the number constraint $N=\int d\textbf{r}\ n\left(\textbf{r}\right)$, which gives
\begin{eqnarray}
e^{\mu_0/k_BT}=N\left(\frac{\lambda\theta\omega\hbar}{k_BT}\right)^3.\label{mu}
\end{eqnarray}

From the normalization condition
\begin{eqnarray}
N=\int d\textbf{r}\int\frac{d\textbf{p}}{\left(2\pi\hbar\right)^3}W\left(\textbf{p},\textbf{r}\right),
\end{eqnarray}
the chemical potential $\mu_0$ is determined by Eq. (\ref{mu}). Under this ansatz, the density distribution in momentum space is given by
\begin{equation}
n\left(\textbf{p}\right)=N\left(\frac{2\pi\hbar^2\theta^2}{mk_BT}\right)^{3/2}\exp\left[ -\frac{\theta^2}{2mk_BT}\left( \frac{p_\rho^2}{\alpha}+\alpha^2p_z^2 \right)\right].\label{MomentumDistribution}
\end{equation}
Thus, the aspect ratio of the momentum space distribution,
which is the ratio of the root-mean-square momentum in the $p_z$ direction to that in a direction
in the $p_x$--$p_y$ plane (we choose the $p_x$ direction in the following) $\sqrt{\langle p_z^2\rangle/\langle p_x^2\rangle}$,
becomes $\alpha^{-3/2}$.
Similarly, we obtain the density distribution in real space as
\begin{equation}
n\left(\textbf{r}\right)=N\left( \frac{2\pi m\omega^2\lambda^2}{k_BT} \right)^{3/2}\exp\left[ -\frac{m\omega^2\lambda^2}{2k_BT}\left(\beta\rho^2+\frac{z^2}{\beta^2} \right)\right],\label{RealDistribution}
\end{equation}
leading to the aspect ratio of the real-space distribution (i.e., the ratio of the root-mean-square radius in the $z$ direction to that in the $x$ direction) being $\sqrt{\langle z^2\rangle/\langle x^2\rangle}=\beta^{3/2}$.
We note that, from Eq. (\ref{RealDistribution}), $\lambda>1$ ($\lambda < 1$) corresponds to compression (expansion) of the gas in real space, which is a consequence of the effective attraction (repulsion) of the dipole--dipole interaction. In addition, comparison of Eq. (\ref{MomentumDistribution}) with Eq. (\ref{RealDistribution}) reveals that $\theta$ plays the same role in real space as $\lambda$ plays in momentum space.

To find a metastable state at a temperature $T$, we look for a local minimum of the Helmholtz free energy
\begin{eqnarray}
F=E-TS.
\end{eqnarray}
Here, the total energy $E$ is the sum of the kinetic energy $E_{K}$, the trapping potential energy $E_{V}$, the Hartree direct energy $E_{H}$, and the Fock exchange energy $E_{ex}$:
\begin{eqnarray}
E&=&E_{K}+E_{V}+E_{H}+E_{ex},
\end{eqnarray}
where the four contributions are given in terms of the Wigner distribution function as~\cite{Miyakawa2008}
\begin{eqnarray}
E_K&=&\frac{1}{2m}\int d\textbf{r}\int \frac{d\textbf{p}}{\left(2\pi\hbar\right)^3}p^2 W\left(\textbf{p},\textbf{r}\right),\label{energy_1}\\
E_V&=&\int d\textbf{r}\int \frac{d\textbf{p}}{\left(2\pi\hbar\right)^3}V\left(\textbf{r}\right) W\left(\textbf{p},\textbf{r}\right),\label{energy_2}\\
E_{H}&=&\frac{1}{2}\int d\textbf{r}\int d\textbf{r}^\prime \int\frac{d\textbf{p}}{\left(2\pi\hbar\right)^3}\int \frac{d\textbf{p}^\prime}{\left(2\pi\hbar\right)^3}V_{dd}\left(\textbf{r}-\textbf{r}^\prime\right) W\left(\textbf{p},\textbf{r}\right) W\left(\textbf{p}^\prime,\textbf{r}^\prime\right),\label{energy_3}\\
E_{ex}&=&-\frac{1}{2}\int d\textbf{R}\int d\textbf{s}\int \frac{d\textbf{p}}{\left(2\pi\hbar\right)^3}\int\frac{d\textbf{p}^\prime}{\left(2\pi\hbar\right)^3}V_{dd}\left(\textbf{s}\right)e^{i\left(\textbf{p}-\textbf{p}^\prime\right)\cdot\textbf{s}/\hbar}W\left(\textbf{p},\textbf{R}\right)W\left(\textbf{p}^\prime,\textbf{s}\right).\label{energy_4}
\end{eqnarray}
In Eq. (\ref{energy_4}), we have introduced the center of mass coordinate $\textbf{R}=\left(\textbf{r}+\textbf{r}^\prime\right)/2$ and the relative coordinate $\textbf{s}=\textbf{r}-\textbf{r}^\prime$. In deriving Eqs. (\ref{energy_3}) and (\ref{energy_4}), we used the mean-field decoupling $\langle \hat\Psi^\dagger\left(\textbf{r}\right)\hat{\Psi}^\dagger\left(\textbf{r}^\prime\right)\hat{\Psi}\left(\textbf{r}^\prime\right)\hat{\Psi}\left(\textbf{r}\right)\rangle\simeq \langle\hat{\Psi}^\dagger\left(\textbf{r}\right)\hat{\Psi}\left(\textbf{r}\right)\rangle \langle\hat{\Psi}^\dagger\left(\textbf{r}^\prime\right)\hat{\Psi}\left(\textbf{r}^\prime\right)\rangle - \langle\hat{\Psi}^\dagger\left(\textbf{r}\right)\hat{\Psi}\left(\textbf{r}^\prime\right)\rangle\langle\hat{\Psi}^\dagger\left(\textbf{r}^\prime\right)\hat{\Psi}\left(\textbf{r}\right)\rangle$. We note that the minus sign in Eq. (\ref{energy_4}) arises from the Fermi statistics. The entropy $S$ is given by
\begin{eqnarray}
S&=&-k_B\int d\textbf{r}\int \frac{d\textbf{p}}{\left(2\pi\hbar\right)^3}\left\{W\left(\textbf{p},\textbf{r}\right)\ln\left[ W\left(\textbf{p},\textbf{r}\right) \right]+\left[1-W\left(\textbf{p},\textbf{r}\right)\right]\ln\left[1-W\left(\textbf{p},\textbf{r}\right)\right]\right\}\nonumber \\
	&\simeq&-k_B\int d\textbf{r}\int\frac{d\textbf{p}}{\left(2\pi\hbar\right)^3}W\left(\textbf{p},\textbf{r}\right)\ln\left[W\left(\textbf{p},\textbf{r}\right)\right].\label{entropy}
\end{eqnarray}
In the above expression for the entropy, we have assumed $W\ll 1$, which is consistent with the Maxwell--Boltzmann distribution.

Using the variational function (\ref{Wigner2}) in Eqs. (\ref{energy_1})--(\ref{entropy}), we can express the free energy in terms of the variational parameters:
\begin{eqnarray}
F&=&\left\{\frac{1}{2\theta^2}f\left(\alpha\right)+\frac{1}{2\lambda^2}g\left(\beta\right)	+\frac{C\lambda^3}{\left(k_BT\right)^{5/2}}\left[I\left(\beta\right)-I\left(\alpha\right)\right]\right.\nonumber \\
	&&\ \ \Bigl.-3+\ln\left[N\left(\lambda\theta\omega\hbar\right)^3\right]-3\ln\left[k_BT\right]\Bigr\}{Nk_BT},\label{FreeEnergy}
\end{eqnarray}
where $C\equiv \left(d^2m^{3/2}\omega^3 N\right)/\left(3\cdot 2^2\pi^{1/2}\right)$. In the above equation, we have defined the scaling functions $f\left(\alpha\right)=2\alpha+1/ \alpha^2$ and $g\left(\beta\right)=2\beta_0/\beta+\beta^2/\beta_0^2$, where $\beta_0\equiv\left(\omega_\rho/\omega_z\right)^{3/2}$. The deformation function $I\left(\alpha\right)$ is defined in Refs. \cite{Miyakawa2008,Sogo2009}. The first and second terms in the first line of Eq. (\ref{FreeEnergy}) are the kinetic energy and the potential energy, respectively. We note that the kinetic energy is inversely proportional to the square of $\theta$. Thus, an increase in $\theta$ indicates a reduction in the effective temperature $T^*=T/\theta^2$ associated with the kinetic energy. The third and fourth terms are the Hartree direct energy and the Fock exchange energy by the dipole--dipole interaction, respectively.
Note that the deformation parameters $\alpha$ and $\beta$ contribute separately to the internal energy, $E$.
The momentum space deformation parameter $\alpha$ appears only in the kinetic and Fock exchange energies,
whereas the real space deformation parameter $\beta$ appears only in the potential and Hartree direct energy.
The last three terms in Eq. (\ref{FreeEnergy}) express the entropy, which is independent of $\alpha$ and $\beta$.

The equilibrium solution can be found by minimizing the free energy $F$ with respect to the four variational parameters,
$\alpha$, $\beta$, $\lambda$, and $\theta$. To be a local minimum of $F$ representing a metastable state, the solution must be a convex downward point in the four-dimensional space of the variational parameters.
This requires the following conditions:
\begin{eqnarray}
\left\{
\begin{array}{l}
0<\alpha<1,\\
\beta_0<\beta,
\end{array}
\right.
\label{analyticcondition}
\end{eqnarray}
From the conditions~(\ref{analyticcondition}), we find that the momentum space distribution is always elongated in the $p_z$ direction The real-space distribution tends to be stretched in the $z$ direction and the aspect ratio of the real-space distribution is larger than that for the non-interacting case.
These results are consistent with those for the zero-temperature case~\cite{Miyakawa2008}.

We find the equilibrium solution by minimizing Eq. (\ref{FreeEnergy}) numerically by varying three system parameters (temperature $T$, trap aspect ratio $\beta_0$, and the magnitude of the electric dipole moment $p$) for $m=100$ a.u.m., $\omega=2\pi\times 10^2 $ Hz, and $N=10^4$.

\begin{figure}
  \begin{center}
      \scalebox{0.5}[0.5]{\includegraphics{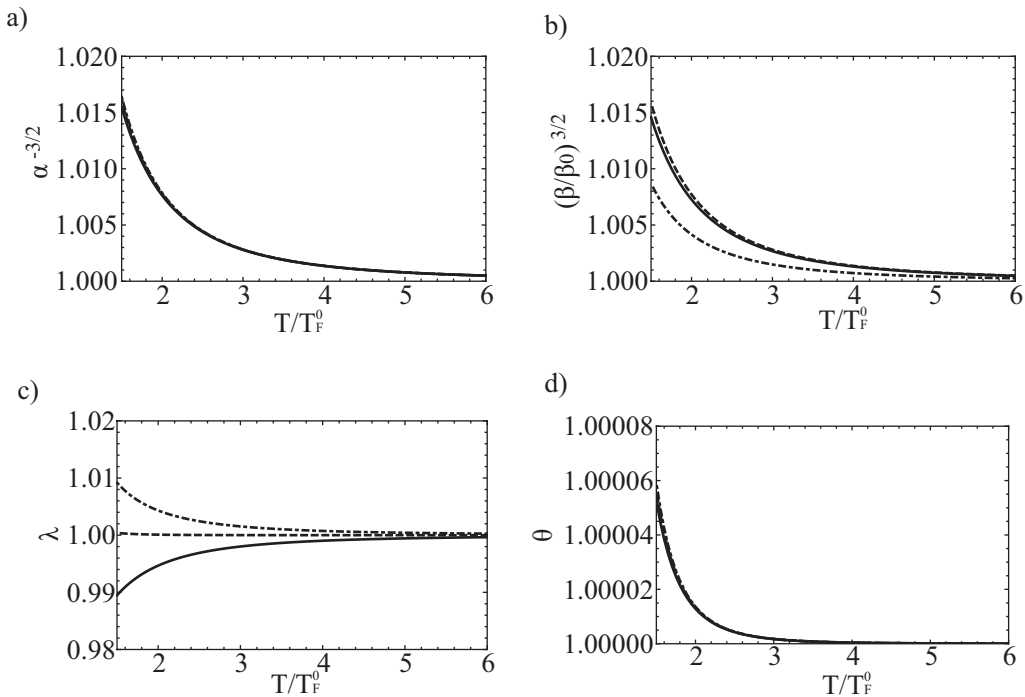}}
       \caption{Variational parameters as functions of the temperature $T$ for a fixed dipole moment $p=1 {\rm \ Debye}$ for a trap asymmetry $\beta_0=0.5$ (solid line), $\beta_0=1.0$ (dashed line), and $\beta_0=2.0$ (dash-dotted line).}
    \label{fig:DFTb0}
  \end{center}
\end{figure}

Figure~\ref{fig:DFTb0} shows a plot of the variational parameters as a function of temperature in units of ideal gas Fermi temperature $T_F^0=\left(6N\right)^{1/3}\hbar\omega\simeq1.88$ nK for $\beta_0=0.5$ (solid line), $\beta_0=1.0$ (dashed line), and $\beta_0=2.0$ (dash-dotted line) and for when the electric dipole moment is fixed at $p=1 {\rm \ Debye}$. Figure~\ref{fig:DFTb0}(a) indicates that the momentum distribution becomes more elongated in the dipolar direction as the temperature decreases. The deformation in the momentum distribution arises from competition between the kinetic energy, which favors an isotropic momentum distribution, and the exchange energy, which favors an anisotropic momentum distribution. As the temperature decreases, the kinetic energy decreases while the interaction energy increases, leading to a large aspect ratio in the momentum distribution.
Figure~\ref{fig:DFTb0}(b) reveals that a lower temperature gives rise to a larger deviation of the aspect ratio in real-space from the non-interacting case. It also shows that the shift of $\beta$ is largest in the isotropic trap.
Figure~\ref{fig:DFTb0}(c) shows that the magnitude of the deviation from $1$ in $\lambda$ is larger at lower temperatures. In addition, we have $\lambda>1$ for $\beta_0>1$ (cigar-shaped trap) and $\lambda<1$ for $\beta_0<1$ (oblate trap).
The Hartree direct energy becomes attractive (repulsive) in the former (latter) case due to deformation of the spatial density distribution mainly determined by the trap geometry, which leads to compression (expansion) of the gas cloud for $\beta_0>1$ ($\beta_0<1$).
Finally, Fig.~\ref{fig:DFTb0}(d) reveals that the dipolar Fermi gas is compressed in momentum space irrespective of the trap aspect ratio because only the kinetic energy and entropy depend on $\theta$ (Eq. (\ref{FreeEnergy})). In addition, the momentum distribution is not compressed when there is no exchange energy. Whereas the interaction effect increases at lower temperatures in all cases,
the deformation in both real and momentum spaces are invisibly small for $T>2T_F^0$ and $p=1\ {\rm Debye}$.

 \begin{figure}
  \begin{center}
      \scalebox{0.5}[0.5]{\includegraphics{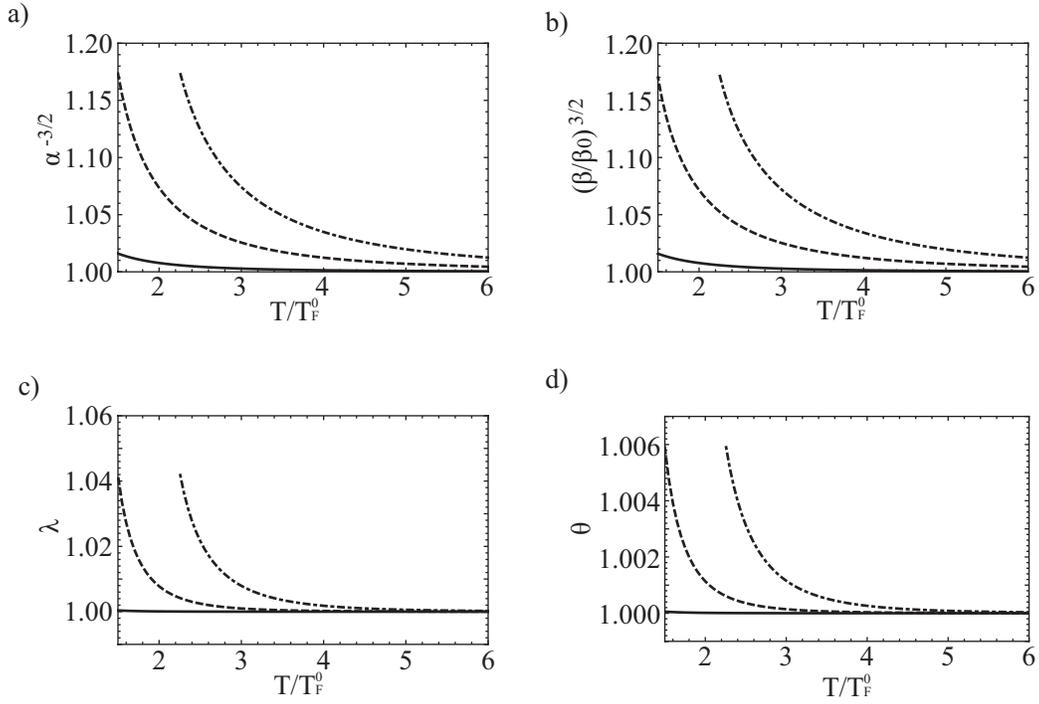}} 
      \caption{Variational parameters as functions of the temperature for a fixed trap asymmetry $\beta_0=1.0$ for a dipole moment $p=1.0 {\rm \ Debye}$ (solid line), $p=3.0 {\rm \ Debye}$ (dashed line), and $p=5.0 {\rm \ Debye}$ (dash-dotted line).}
    \label{fig:DFTp}
  \end{center}
\end{figure}

Figure~\ref{fig:DFTp} shows the variational parameters as a function of temperature for $p=1.0$ Debye (solid line),
$p=3.0$ Debye (dashed line), and $p=5.0$ Debye (dash-dotted line) for an isotropic trap $\beta_0=1$.
In all cases, the deviation of the variational parameters from those of the non-interacting case becomes more pronounced at lower temperatures and at higher electric dipole moments.
For large electric dipole moments, we plotted the optimized values for $T>T_c$,
where $T_c$ is the critical temperature below which the variational free energy~(\ref{FreeEnergy}) has no local minimum and the system becomes unstable to collapse. The critical temperature for this collapse for $p=5.0$ Debye is $T_c={2.25}T_F^0$. On the other hand, systems with $p=1.0$ Debye and $p=3.0$ Debye are stable in the temperature regime $T>1.5T_F^0$. These results reveal that the system is unstable even in the high-temperature regime for large electric dipole moments. Figure~\ref{fig:DFTp}(a) and (b) shows that the aspect ratios in both momentum and real space are sufficiently large to observe deformation effects in experiments for polar molecules with large electric dipole moments.

 \begin{figure}
  \begin{center}
      \scalebox{0.5}[0.5]{\includegraphics{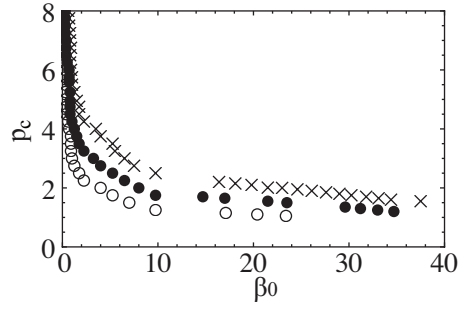}} 
    \caption{Critical value of the dipole moment $p_c$ as a function of the trap aspect ratio for
    $T/T_F^0=1.5$ (open circles), $T/T_F^0=2.0$ (filled circles), and $T/T_F^0=3.0$ (crosses).}
  \label{fig:DF_instability_1}
  \end{center}
  \end{figure}

\begin{figure}
  \begin{center}
      \scalebox{0.5}[0.5]{\includegraphics{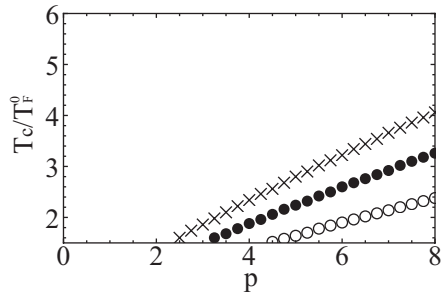}} 
    \caption{Critical temperature as a function of the dipole moment for trap aspect ratios $\beta_0={\rm 0.5}$ (open circles), $\beta_0={\rm 1.0}$ (filled circles), and $\beta_0={\rm 2.0}$ (crosses).}
  \label{fig:DF_instability_2}
  \end{center} 
  \end{figure}

\begin{figure}
  \begin{center}
      \scalebox{0.5}[0.5]{\includegraphics{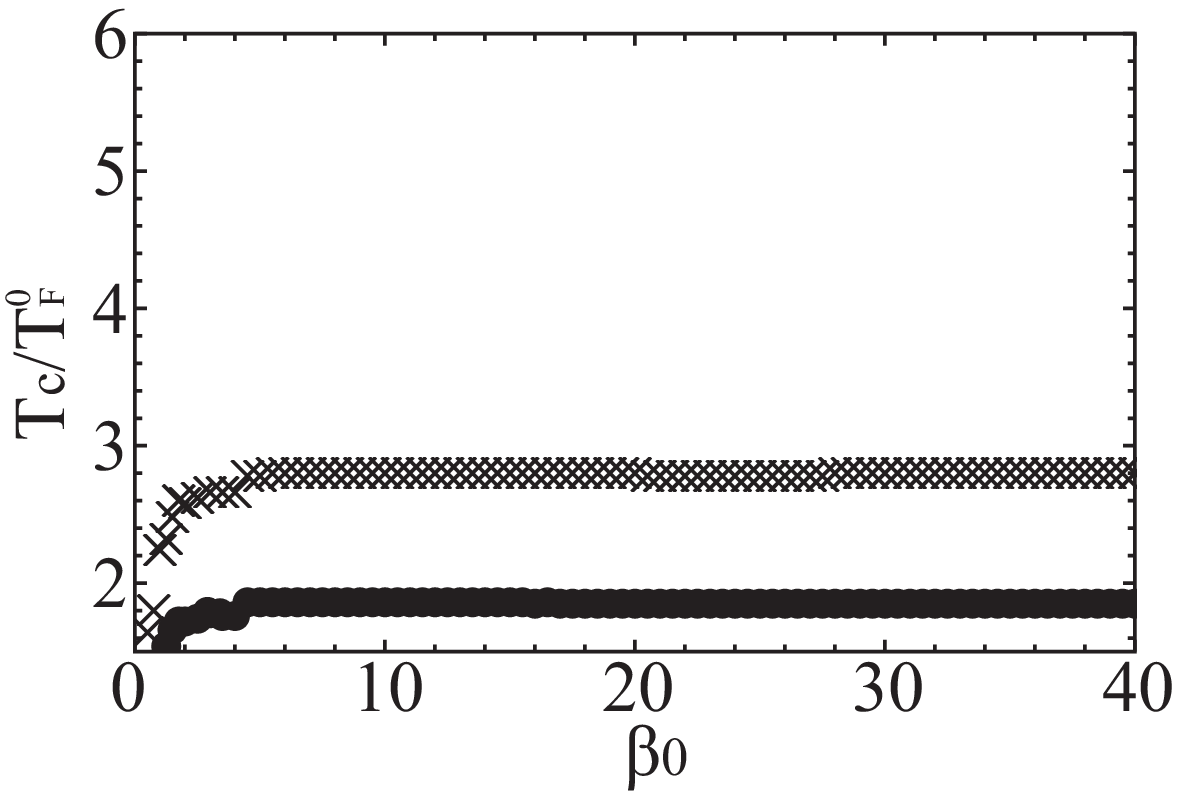}} 
    \caption{Critical temperature as a function of the trap aspect ratio for dipole moments $p={\rm 3.0}$ (filled circles) and $p={\rm 5.0}$ (crosses).}
  \label{fig:DF_instability_3}
  \end{center}
\end{figure}

Figure~\ref{fig:DF_instability_1} shows the stability diagram as functions of the electric dipole moment $p$ and trap aspect ratio $\beta_0$ for different temperatures.
This result indicates that the critical dipole moment $p_c$ increases drastically as the trap becomes more oblate. This is consistent with the result for the zero-temperature case~\cite{Miyakawa2008}. The unstable region expands with decreasing temperature. 
From Fig.~\ref{fig:DFTp}, we presume that the critical temperature for collapse instability increases as the electric dipole moment increases. This behavior is confirmed by Fig.~\ref{fig:DF_instability_2}, which shows the critical temperature $T_c$ for $\beta_0=0.5$, $\beta_0=1.0$, and $\beta_0=2.0$.
Figure~\ref{fig:DF_instability_3} shows the critical temperature as a function of the trap aspect ratio for $p=1.0$ Debye and $p=2.0$ Debye. This result indicates that the critical temperature $T_c$ increases more as $\beta_0$ increases.
When ${\it p}{\rm =1.0 \ Debye}$, the system is always stable within the parameter range shown in the figure ($\beta_0 \leq 40$). Figure~\ref{fig:DF_instability_3} suggests that we should be careful not to enter the unstable region when the temperature is reduced. Even if the system is initially in a stable region at a high temperature, the system may (depending on the trap aspect ratio) become unstable on cooling before reaching the quantum degenerate regime. In addition, we see that the local minimum always disappears when $\lambda$ exceeds a critical value. This suggests that the instability first occurs in real space and a dipolar Fermi gas collapses when $\lambda$ reaches the critical value.

Finally, we discuss the stability properties in connection with experiments. The JILA experiment~\cite{Ni2008} revealed that the mass of the polar molecule is about ${\it m}{\rm=128 \ a.m.u.}$ and that the electric dipole moment is about ${\it p}{\rm=0.566 \ Debye}$. Taking these values along with $\omega=2\pi\times 10^2$ and $N=10^4$, it is always possible to find a stable dipolar Fermi gas for any temperature $T\geq T_F^0$ in the trap aspect ratio regime $\beta_0\leq 40$.
Since, according to the results in Ref. \cite{Miyakawa2008}, a gas of polar molecules at zero temperature with the same system parameters becomes unstable, the critical temperature should lie in $0< T <T_F^0$. However, the dipolar Fermi gas becomes unstable at extremely large trap aspect ratios. For $\omega=2\pi\times10^2$, $N=10^4$, $p=0.566$ Debye, and $m=128$ {a.m.u.}, the critical trap aspect ratios for $T=1.5T_F^0$ and $T=2.0T_F^0$ are $\beta_0^c=62.0$ and $\beta_0^c=72.5$, respectively.

We discuss how the deformation effects of momentum-space and real-space distributions can be observed in the polar molecules used in the JILA experiment~\cite{Ni2008}.
For the same parameters given in the previous paragraph except for the frequency of the harmonic oscillator potential $\omega$, increasing $\omega$ causes the dipole--dipole interaction effects to become very pronounced.
This is because the density of the gas trapped in a tight potential becomes high and the interaction becomes more effective. For $\omega=5\times2\pi \times 10^2$ Hz, $\beta_0{\rm =1.0}$, and $T=1.5T_F^0$, we have aspect ratios $\sqrt{\langle p_z^2\rangle/\langle p_x^2\rangle}=\alpha^{-3/2}\simeq 1.130$ in the momentum distribution and $\sqrt{\langle z^2\rangle/\langle x^2\rangle}=\beta^{3/2}\simeq 1.127$ in the spatial distribution. For $\omega=6\times2\pi \times 10^2$ Hz, $\beta_0{\rm =1.0}$ and $T=2.0T_F^0$, we have $\alpha^{-3/2}\simeq 1.106$ and $\beta^{3/2}\simeq 1.103$. Thus, the deformation effects arising from the anisotropic nature of the dipole--dipole interaction are detectable in the present experimental conditions.

In conclusion, we have used a variational method to study the equilibrium properties of a dipolar Fermi gas at finite temperatures. As at zero temperature, the anisotropic nature of the dipole--dipole interaction leads to deformations in momentum and real space and the partial attraction of the interaction causes instability of the gas to collapse. In addition, we found that the dipolar Fermi gas becomes compressed in momentum space as the electric dipole moment increases. We found that the deformations in both momentum and real space can be observed in the high-temperature regime with a large electric dipole moment and a high trap frequency. In addition, we found that the stable region expands at finite temperatures. These results will be useful when cooling polar molecules in experiments. We hope this study will stimulate further experiments on dipolar Fermi gas at finite temperatures.

In future studies, we intend to study dynamics of a dipolar Fermi gas at finite temperatures, such as the expansion dynamics and collective oscillations. In particular, expansion dynamics is important since it can provide direct information of the deformation of momentum distribution. Sogo et al. have studied the expansion dynamics at zero temperature and found that the dipolar Fermi gas expands along the dipole moment direction irrespective of the trap aspect ratio~\cite{Sogo2009}. We will study the temperature dependence of the expansion dynamics in the presence of the dipole--dipole interaction.

{\it Note added. }Recently, we became aware of two works of equilibrium properties of dipolar Fermi gas at finite temperatures~\cite{Zhang2010,Kestner2010}.


\begin{thebibliography}{10}

\bibitem{Griesmaier2005}
Axel Griesmaier, J{\"o}rg Werner, Sven Hensler, J{\"u}rgen Stuhler, and Tilman
  Pfau, Phys. Rev. Lett. {\bf 94},  160401  (2005).

\bibitem{Stuhler2005}
J. Stuhler, A. Griesmaier, T. Koch, M. Fattori, T. Pfau, S. Giovanazzi, P.
  Pedri, and L. Santos, Phys. Rev. Lett. {\bf 95},  150406  (2005).

\bibitem{Giovanazzi2006}
S. Giovanazzi, P. Pedri, L. Santos, A. Griesmaier, M. Fattori, T. Koch, J.
  Stuhler, and T. Pfau, Phys. Rev. A {\bf 74},  013621  (2006).

\bibitem{Santos2000}
L. Santos, G.~V. Shlyapnikov, P. Zoller, and M. Lewenstein, Phys. Rev. Lett.
  {\bf 85},  1791  (2000).

\bibitem{Yi2000}
S. Yi and L. You, Phys. Rev. A {\bf 61},  041604(R)  (2000).

\bibitem{Yi2001}
S. Yi and L. You, Phys. Rev. A {\bf 63},  053607  (2001).

\bibitem{Goral2002_2}
K. G{\'o}ral and L. Santos, Phys. Rev. A {\bf 66},  023613  (2002).

\bibitem{Goral2002}
K. G{\'o}ral, L. Santos, and M. Lewenstein, Phys. Rev. Lett. {\bf 88},  170406
  (2002).

\bibitem{Danshita2008}
Ippei Danshita and Carlos A. R.~{\'S}a de~Melo, arXiv {\bf 0804},  0494
  (2008).

\bibitem{Goral2001}
Krzysztof G{\'o}ral, Berthold-Georg Englert, and Kazimierz Rz{\c{a}}{\.z}ewski,
  Phys. Rev. A {\bf 63},  033606  (2001).

\bibitem{Miyakawa2008}
Takahiko Miyakawa, Takaaki Sogo, and Han Pu, Phys. Rev. A {\bf 77},  061603(R) 
  (2008).

\bibitem{Zhang2009}
J.~N. Zhang and S. Yi, Phys. Rev. A {\bf 80},  053614  (2009).

\bibitem{He2008}
L. He, J.~N. Zhang, Yunbo Zhang, and S. Yi, Phys. Rev. A {\bf 77},  031605(R)
  (2008).

\bibitem{Nishimura2009}
Takushi Nishimura and Tomoyuki Maruyama, arXiv {\bf 0907},  1757  (2009).

\bibitem{Sogo2009}
T. Sogo, L. He, T. Miyakawa, S. Yi, H. Lu, and H. Pu, New. J. Phys. {\bf 11},
  055017  (2009).

\bibitem{Baranov2002}
M. Baranov, L. Dobrek, K. G{\'o}ral, L. Santos, and M. Lewenstein, Phys. Scr.
  {\bf T102},  74  (2002).

\bibitem{Baranov2004}
M.~A. Baranov, {\L}. Dobrek, and M. Lewenstein, Phys. Rev. Lett. {\bf 92},
  250403  (2004).

\bibitem{Bruun2008}
G.~M. Bruun and E. Taylor, Phys. Rev. Lett. {\bf 101},  245301  (2008).

\bibitem{Zhao2009}
Cheng Zhao, Lei Jiang, Xunxu Liu, W.~M. Liu, Xubo Zou, and Han Pu, arXiv {\bf
  0910},  4775  (2009).

\bibitem{Zhang2010}
J.~N. Zhang and S. Yi, arXiv {\bf 1001},  0426  (2010).

\bibitem{Kestner2010}
J.~P. Kestner and S.~Das Sarma, arXiv {\bf 1001},  4763  (2010).

\bibitem{Ni2008}
K.~K. Ni, S. Ospelkaus, M.~H.~G. de~Miranda, A. Pe'er, B. Neyenhuis, J.~J.
  Zirbel, S. Kotochigova, P.~S. Julienne, D.~S. Jin, and J. Ye, Science {\bf
  322},  231  (2008).

\end{thebibliography}

\end{document}